\documentclass[12pt]{iopart}

\usepackage{graphicx}
 \usepackage{dcolumn}
 \usepackage{bm}
 \usepackage{amssymb}
 \usepackage{amsfonts}
 \usepackage{color}
 \usepackage{natbib}
 \usepackage{url}

\newcommand{\refsec}[1]{\ref{sec:#1}}

\graphicspath{{FIGS/}}

\begin{document}

\title{Non-linear modeling of the plasma response to RMPs in ASDEX Upgrade}

\author{F.Orain, M.H\"olzl, E.Viezzer, M.Dunne, M.Willensdorfer, W.Suttrop, E.Strumberger, S.G\"unter, A.Lessig, the ASDEX Upgrade Team}
\address{ Max-Planck-Institut f\"ur Plasmaphysik, D-85748 Garching, Germany}

\author{M.B\'ecoulet, G.T.A.Huijsmans, J.Morales}
\address{CEA-IRFM, Cadarache, 13108 Saint-Paul-Lez-Durance, France}

\author{A.Kirk, S.Pamela}
\address{CCFE, Culham Science Centre, Oxon, OX14 3DB, UK}
\author{P.Cahyna}
\address{Institute of Plasma Physics CAS, Za Slovankou 1782/3, 182 00 Prague 8, Czech Republic}
\author{the EUROfusion MST1 Team}
\address{List of contributors on http://www.euro-fusionscipub.org/mst1}

\ead{forain@ipp.mpg.de}

 \begin{abstract}
The plasma response to Resonant Magnetic Perturbations (RMPs) in ASDEX Upgrade is modeled with the non-linear resistive MHD code JOREK, using input profiles that match those of the experiments as closely as possible. The RMP configuration for which Edge Localized Modes are best mitigated in experiments is related to the largest edge kink response observed near the X-point in modeling. On the edge resonant surfaces $q=m/n$, the coupling between the $m+2$ kink component and the $m$ resonant component is found to induce the amplification of the resonant magnetic perturbation. The ergodicity and the 3D-displacement near the X-point induced by the resonant amplification can only partly explain the density pumpout observed in experiments. 
 \end{abstract}

 \maketitle


\section{Introduction}
The Edge Localized Modes (ELMs), occurring at the plasma edge in tokamaks, induce large transient heat loads on plasma facing components. In ITER, the heat loads on the divertor due to ELMs are expected to be intolerable for materials if unmitigated, hence motivating a broad effort on developing and understanding reliable ways to mitigate the ELMs. One of the promising methods to control ELMs is the application of non-axisymmetric resonant or non-resonant magnetic perturbations (RMPs or NRMPs) by dedicated coils. The original idea is that a small applied magnetic perturbation ($\delta B / B_{toroidal} \sim 10^{-4}$) is expected to induce magnetic islands on the resonant surfaces characterized by the safety factor $q=m/n$ (where $m$ and $n$ are respectively the poloidal and toroidal mode numbers, with the dominant $n$ fixed by the RMP configuration). At the plasma edge where rational surfaces are close to each other, consecutive island chains are likely to overlap and induce an ergodic layer \citep{Ghendrih_PPCF96}. The radial particle and heat transport being enhanced in the ergodic zone, RMPs should be able to slightly deconfine the plasma edge and reduce the pedestal pressure gradient under the ELM-triggering threshold (so far understood in the framework of the Peeling-Ballooning or P-B theory) \cite{Snyder_PoP02, Liang_15_book_ELMs}. In addition, RMPs are also believed to lower the P-B stability boundary limit, resulting in more frequent, smaller ELMs \citep{Chapman_PoP13, Ham_PPCF15}.

When applying RMPs, the mitigation or suppression of type-I ELMs was successfully obtained in DIII-D, ASDEX Upgrade (AUG), JET, MAST, KSTAR, NSTX and EAST tokamaks \cite{Evans_PRL, Suttrop_PRL11, Liang_PRL07, Kirk_NF10, Jeon_PRL12, Canik_NF10, Wan_NF15}. However the conditions to obtain a strong mitigation or complete ELM suppression have been proven to be more complicated than this simple picture, due to the strong screening of RMPs by plasma flows. Even though the understanding of the plasma response to RMPs has been much improved in recent years \citep{Strauss_NF09, Nardon_NF10, Liu_PoP10, Yu_NF11, Becoulet_NF12, Ferraro_PoP12, Schmitz_JNM13, Orain_PoP13, Becoulet_PRL14, Orain_PPCF15, Liu_NF15, Ryan_PPCF15}, the interaction between plasma and RMPs has to be further studied to better interpret the current experiments and to be capable to predict the effect of RMPs in ITER. In particular, in addition to the resonant response, the role of the so-called ``edge kink response'' or ``peeling response'' has to be assessed, where the amplification of the external field perturbation is observed, probably resulting from the interaction between the applied perturbation and a marginally stable kink mode \citep{Boozer_PRL01}. At low collisionality, the ELM suppression in DIII-D and the strongest ELM mitigation in AUG were recently shown to be related to the excitation of the edge kink (or peeling) response \citep{Paz-Soldan_PRL15, Nazikian_PRL15, Kirk_NF15}.

This paper aims at better understanding the role of both the resonant and the kink responses on the ELM mitigation by RMPs in AUG at low collisionality. In this respect, the plasma response to RMPs was modeled with the non-linear resistive MHD code JOREK \citep{Huysmans_NF07, Czarny_JCP08}, using the data extracted from different RMP configurations tested in ASDEX Upgrade experiments. In Section \refsec{part2}, the experimental discharges considered and the comparison between the experimental data and the kinetic and rotation profiles in simulation are described. In Section \refsec{part3}, the generic features of the plasma response to RMPs are discussed. In Section \refsec{part4}, the impact of the differential phase between upper and lower RMP coils (varied in experiments and simulations accordingly) on the plasma response is assessed. The evolution of the pedestal profiles is then described in Section \refsec{part5}, and discussions and conclusions are provided in Section \refsec{part6}.

\section{Input parameters and comparison to experiments} \label{sec:part2}

\subsection{Experimental discharges}
\begin{figure}[h!]
\centering
\includegraphics[width=0.5\textwidth]{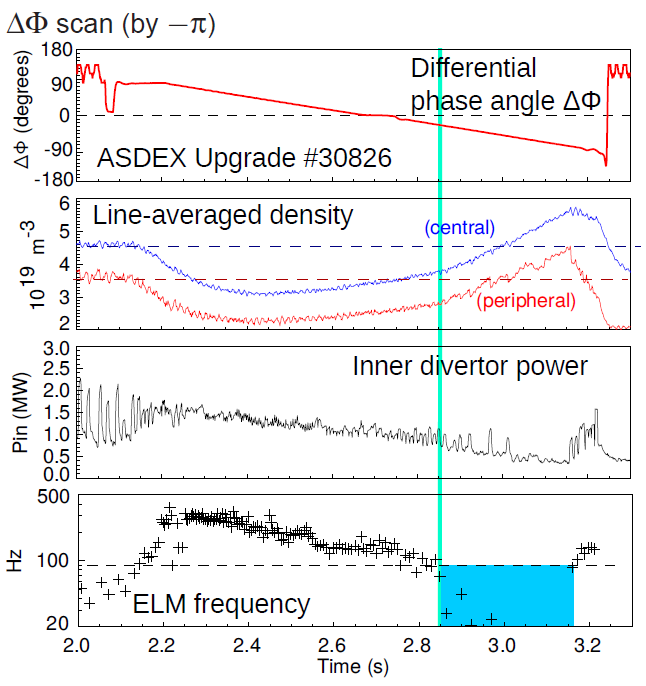}
\caption{AUG shot \#30826 during which the differential phase $\Delta \Phi$ between upper and lower RMP coils is scanned. Time evolution of $\Delta \Phi$, electron density, inner divertor power and ELM frequency is plotted. The strongest ELM mitigation (largest ELM frequency) is obtained for $\Delta \Phi$ between $+90^{\circ}$ and $+60^{\circ}$, and a transient ELM-free phase (in blue) is observed for $\Delta \Phi$ between $-30^{\circ}$ and $-90^{\circ}$.} 
\label{fig:phase_scan}
\end{figure}

The data used in modeling are extracted from AUG shots \#31128 and \#30826. In the discharge \#30826 presented in Fig.\ref{fig:phase_scan}, the differential phase between the magnetic field perturbation generated by upper and lower RMP-coils \citep{Suttrop_PPCF_2011, Suttrop_IAEA14} is slowly rotated between $\Delta \Phi= +90^{\circ}$ and $\Delta \Phi= -90^{\circ}$ by varying the current applied in the coils. In this shot, the strongest ELM mitigation (characterized by the largest ELM frequency, see Fig.\ref{fig:phase_scan}) is obtained for  $\Delta \Phi$ between $+90^{\circ}$ and $+60^{\circ}$ and it is correlated with the strongest density pumpout observed. The increase in ELM frequency is probably induced by the reduction of the pedestal stability limit due to the application of RMPs \citep{Chapman_PoP13, Ham_PPCF15}. Note that a transient ELM-free phase (in blue) is observed for $\Delta \Phi= -30^{\circ}$ to $-90^{\circ}$: this phenomenon is not fully understood, but a possible explanation is that the stability limit may be enhanced in these RMP configurations, allowing the pedestal to grow growing until it reaches the new stability limit \citep{Suttrop_IAEA14}. This ELM-free state was only obtained in transient phases so far. In our simulations, steady magnetic perturbations are considered for $\Delta \Phi= +90^{\circ}$ (corresponding to the shot \#31128), $+60^{\circ}$, $+30^{\circ}$, $0^{\circ}$, $-30^{\circ}$, $-60^{\circ}$ and $-90^{\circ}$).

\begin{figure}[h!]
\begin{minipage}[]{0.4\linewidth}
\centering
\includegraphics[width=1.08\columnwidth]{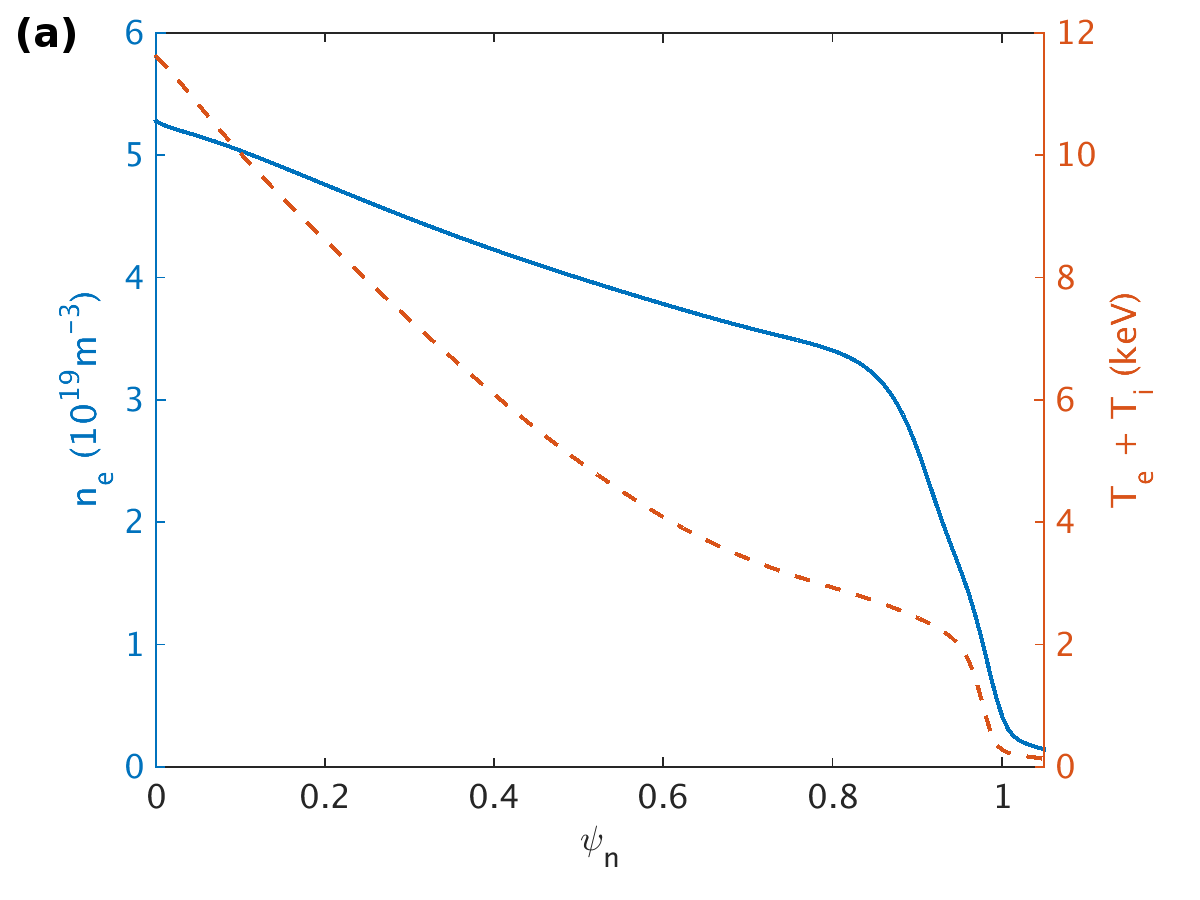}
\end{minipage}
\begin{minipage}[]{0.4\linewidth}
\centering
\vspace{-0.3cm}
\includegraphics[width=1.08\columnwidth]{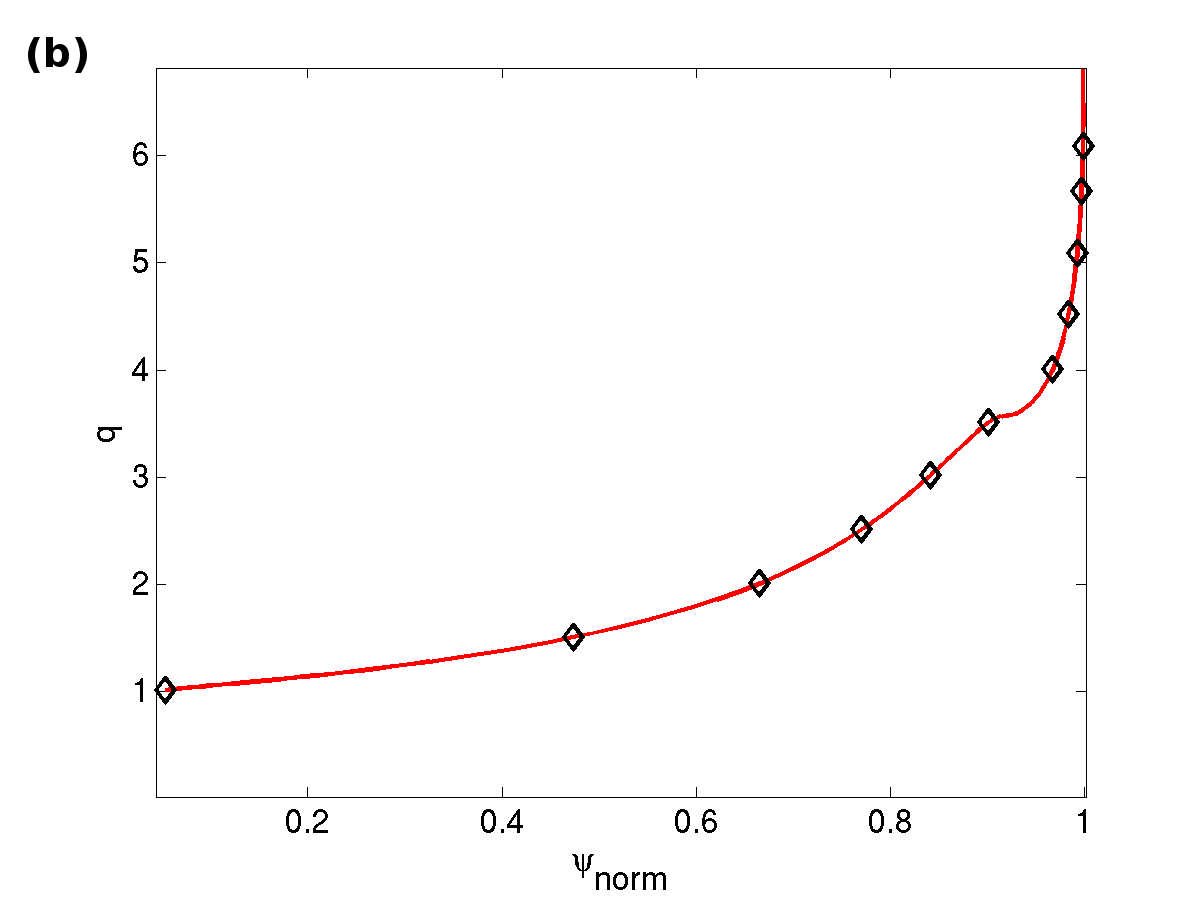} 
\end{minipage}
\centering
\caption{Radial profiles of (a) the electron density (blue line) and total (ion+electron) temperature (red dashed line). (b) safety factor q profile with positions of resonant surfaces (black diamonds).}
\label{fig:input_profiles}
\end{figure}

Initial electron density $n_e$ and total (electron+ion) temperature $T$ profiles (plotted in Fig.\ref{fig:input_profiles} (a) as a function of the normalized poloidal flux $\psi_n$) are fitted from experimental data with pedestal gradients that are enhanced within the error bars in order to make the plasma peeling-ballooning unstable. The equilibrium reconstruction and the $FF'$ profile calculation are made with the equilibrium code CLISTE \cite{CLISTE}. The resulting q-profile is plotted in Fig.\ref{fig:input_profiles}(b), with the position of the resonant surfaces $q=m/2$ (for an $n=2$ perturbation) marked as black diamonds.

The JOREK extended reduced MHD model described in Ref.\cite{Orain_PoP13} is used for modeling. The equations are solved for the following variables: magnetic flux $\psi$, mass density $\rho=m_i n_e$ (with $m_i$ the ion mass and $n_e$ the electron density), temperature $T$, parallel velocity $V_{||}$, electric potential $u$, toroidal current $j=\Delta*\psi$ and vorticity $W=\nabla_\perp^2u$. 
Perpendicular transport is modeled with diffusive heat and particle terms (with heat and particle diffusivity coefficients reduced in the pedestal to reproduce the Edge Transport Barrier) balanced with heat and particle sources, such that the temperature and density profiles do not evolve in time in axisymmetric simulations ($n=0$ only). For numerical reasons, the central resistivity is taken $\sim 10$ times larger than the Spitzer value and the resistivity profile follows a $T^{-3/2}$ dependence.

A simulation is run as follows: in a first step, the axisymmetric equilibrium ($n=0$ only) is calculated, and the ($n=2$) perturbations are added in a second step. The axisymmetric equilibrium is strongly influenced by plasma flows. As described in Ref.\cite{Orain_PoP13}, a source of parallel rotation mimicking the experimental rotation profile, the two-fluid diamagnetic flows as well as the neoclassical friction constraining the poloidal velocity are included in the model to self-consistently describe the plasma flows. 

\begin{figure}[h!]
\begin{minipage}[]{0.3\linewidth}
\hspace{-1.5cm}
\includegraphics[width=1.15\columnwidth]{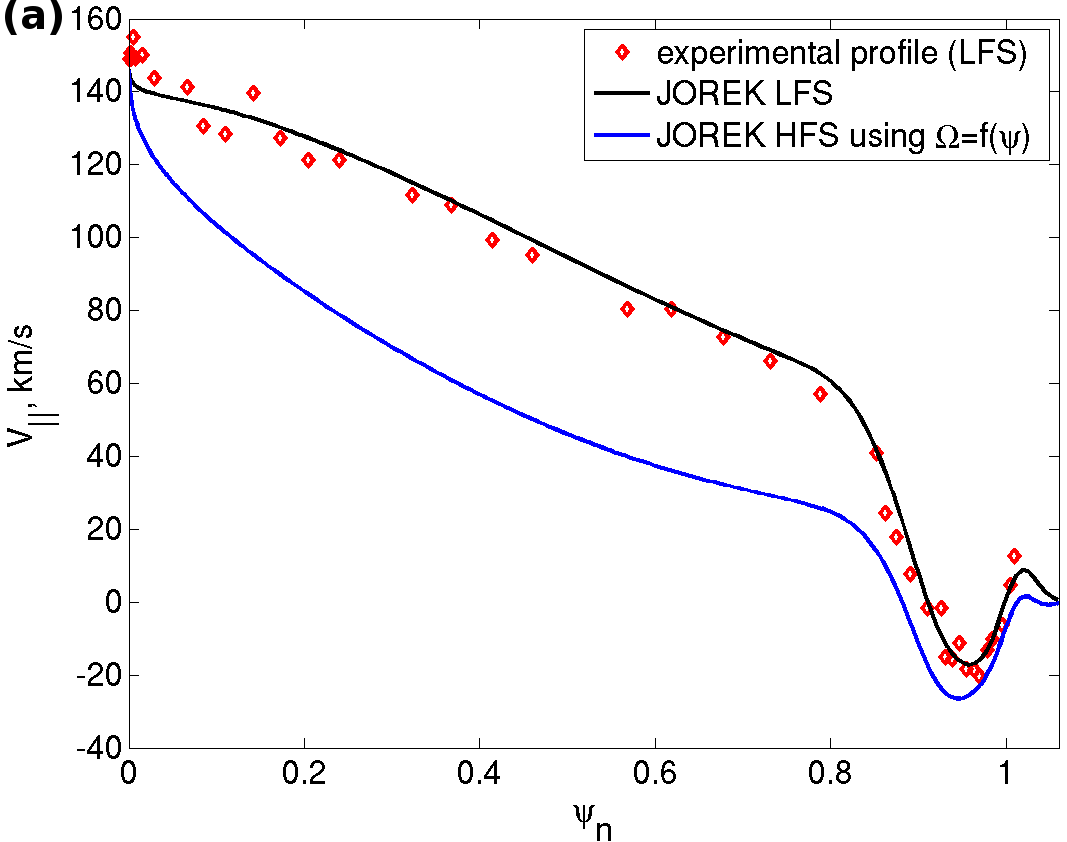}
\end{minipage}
\begin{minipage}[]{0.33\linewidth}
\hspace{-0.5cm}
\includegraphics[width=1.05\columnwidth]{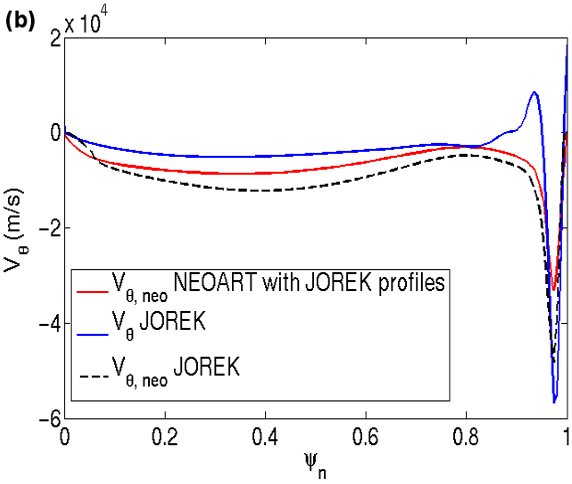} 
\end{minipage}
\begin{minipage}[]{0.3\linewidth}
\hspace{3.5cm}
\includegraphics[width=1.128\columnwidth]{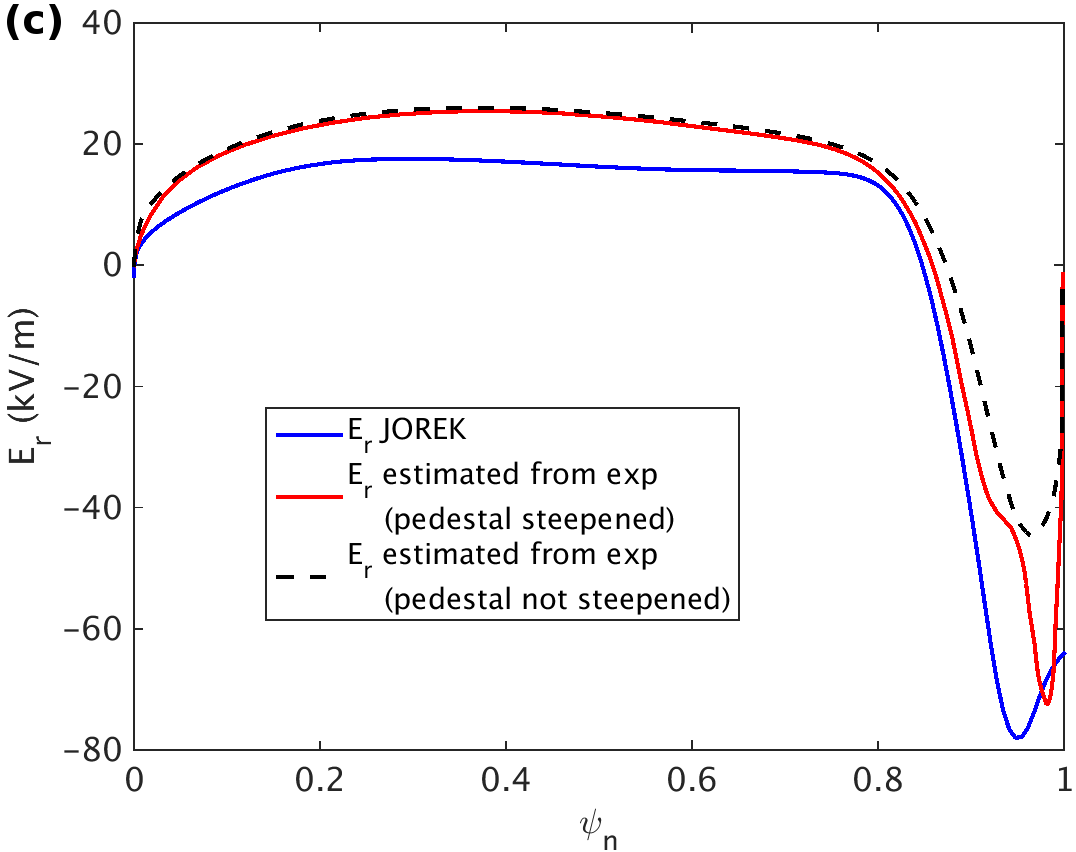} 
\end{minipage}
\centering
\caption{Radial profiles of: (a) Parallel velocity in JOREK at the low field side midplane (black) with comparison to the experimental profile (red squares), and at the high field side midplane (blue). (b) Neoclassical poloidal velocity calculated in NEOART (red) and JOREK (black dashed line), and poloidal velocity in JOREK. (c) Radial electric field $E_r$ in JOREK (blue), $E_r$ estimation from experimental data using the NEOART neoclassical poloidal velocity with steepened pedestal (red) and with pedestal not steepened (black dashed line).}
\label{fig:equil_flows}
\end{figure}

For the source of parallel rotation (considered to be close to the toroidal rotation), the (main) ion toroidal rotation frequency $\Omega_{tor}$ was extracted from experimental measurements at the outboard midplane, assuming that main ions and impurity ions rotate similarly in toroidal direction. The toroidal rotation frequency was taken as a flux function ($\Omega_{tor} = f(\psi)$), consistent with the experimental observations to lowest order \citep{Viezzer_PPCF13}. The source of toroidal rotation was then implemented as a constraint in the viscous term of the parallel momentum equation: $\mu \Delta(V_{||} - \Omega_{tor} \cdot 2\pi R$), where $\mu$ denotes the viscosity coefficient. This way, the equilibrium parallel rotation profile (plotted in Fig.\ref{fig:equil_flows}(a)) perfectly matches the experimental profile at the outboard midplane. The parallel rotation profile at the inboard midplane, corresponding to $V_{||} =\Omega_{tor}(\psi). 2\pi R$, is also given.

As for the main ion poloidal rotation, it cannot be directly extracted from experiments, but previous studies showed that the poloidal rotation is very close to the neoclassical prediction in the pedestal \citep{Viezzer_NF13}, with a good match between experimental data and calculations with the NEOART code \citep{Viezzer_NF13, NEOART_Peeters_PoP00}. The neoclassical poloidal velocity calculated with JOREK (Fig.\ref{fig:equil_flows}(b)) shows good agreement with the NEOART calculation using the same steepened temperature and density profiles as input. The JOREK value (black dashed line) is slightly larger than the NEOART value in the pedestal due to the effective charge $Z_{eff}$ taken as 1 in JOREK. Note that both JOREK and NEOART poloidal neoclassical velocities would have been smaller if the temperature and density profiles had not been steepened, thus the neoclassical poloidal rotation is possibly overestimated in JOREK. The neoclassical friction is implemented in JOREK as a neoclassical tensor $\nabla \cdot \Pi_{i,neo} \propto \mu_{neo} (V_\theta - V_{\theta,neo})$ \citep{Orain_PoP13}. The resulting neoclassical velocity in JOREK $V_\theta$ (plotted in blue) is close to $V_{\theta,neo}$ but slightly differs due to other terms (among others the stress tensor) constraining the velocity in the poloidal momentum equation.

The resulting equilibrium radial electric field $E_r$, calculated from the equilibrium force balance, is given in Fig.\ref{fig:equil_flows}(c). The $E_r$ radial profile in JOREK matches rather well the $E_r$ estimated from experimental profiles using the NEOART calculation of the neoclassical poloidal velocity. Note that the $E_r$ value estimated from experimental profiles would be smaller by a factor of two in the pedestal if the pedestal density and temperature profiles were not steepened to make the plasma P-B unstable (Fig.\ref{fig:equil_flows}(c), black dashed line). In both cases, this represents a large radial electric field well in the pedestal (and thus a large electron perpendicular rotation), capable to strongly screen RMPs at the pedestal top, as discussed below.

\section{Generic features of the plasma response to RMPs} \label{sec:part3}

\begin{figure}[h!]
\centering
\includegraphics[width=1.\textwidth]{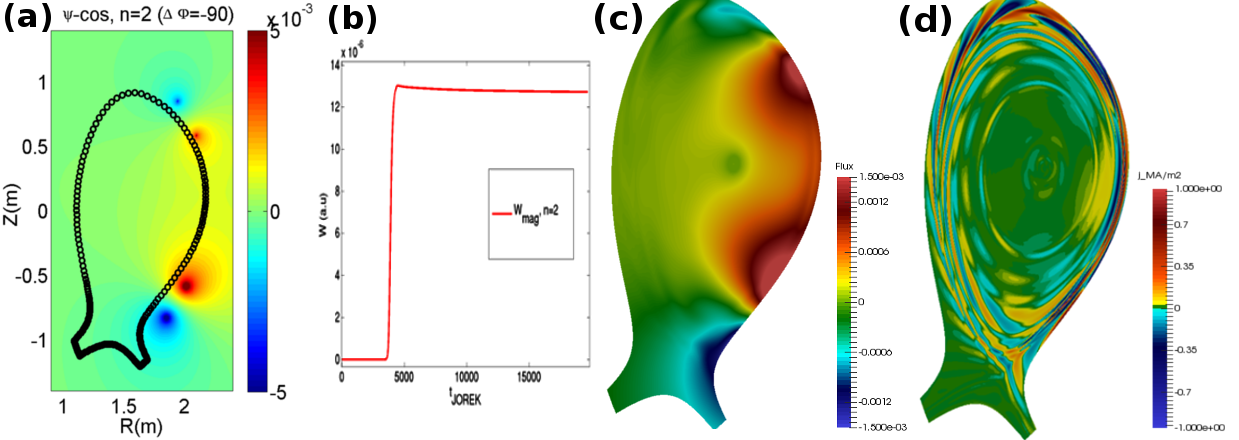} 
\centering
\caption{(a) Vacuum magnetic flux perturbation calculated with the VACFIELD code, and position of the JOREK boundary (black dots) where the vacuum magnetic flux perturbation is applied as boundary condition. (b) Time evolution of the $n=2$ magnetic energy. (c) Poloidal section of the $n=2$ magnetic flux perturbation. (d)  Poloidal section of the $n=2$ current perturbation induced as response to RMPs.}
\label{fig:generic_response}
\end{figure}

Once equilibrium ($n=0$) flows are established, $n=2$ perturbations are added in the simulation. The $n=2$ magnetic flux perturbation induced by RMP coils is beforehand calculated in the vacuum with the VACFIELD code \citep{VACFIELD}, and applied in JOREK as boundary condition for the $n=2$ magnetic flux perturbation. The vacuum field and the JOREK boundary are presented in Fig.\ref{fig:generic_response}(a). The magnetic perturbation at the boundary is progressively increased to its  nominal value in a thousand Alfv\'en times, thus RMPs progressively penetrate into the plasma taking into account the plasma response. The magnetic energy of the $n=2$ mode, plotted in Fig.\ref{fig:generic_response}(b), is growing due to the external RMP application, until it saturates. The 2D-plot of the $n=2$ magnetic flux perturbation penetrating in the plasma (in the $\Delta \Phi= -90^{\circ}$ case) as well as the $n=2$ response toroidal current perturbation on resonant surfaces and in the SOL, are given in Figs.\ref{fig:generic_response}(c-d). 


\begin{figure}[h!]
\begin{minipage}[]{0.4\linewidth}
\centering
\includegraphics[width=1.\columnwidth]{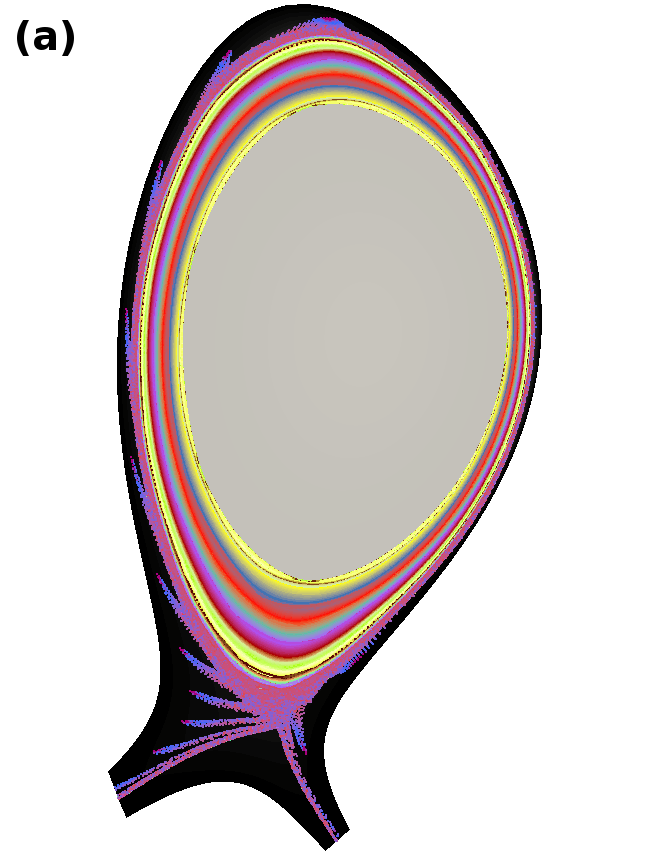}
\end{minipage}
\begin{minipage}[]{0.48\linewidth}
\centering
\vspace{-0.3cm}
\includegraphics[width=1.\columnwidth]{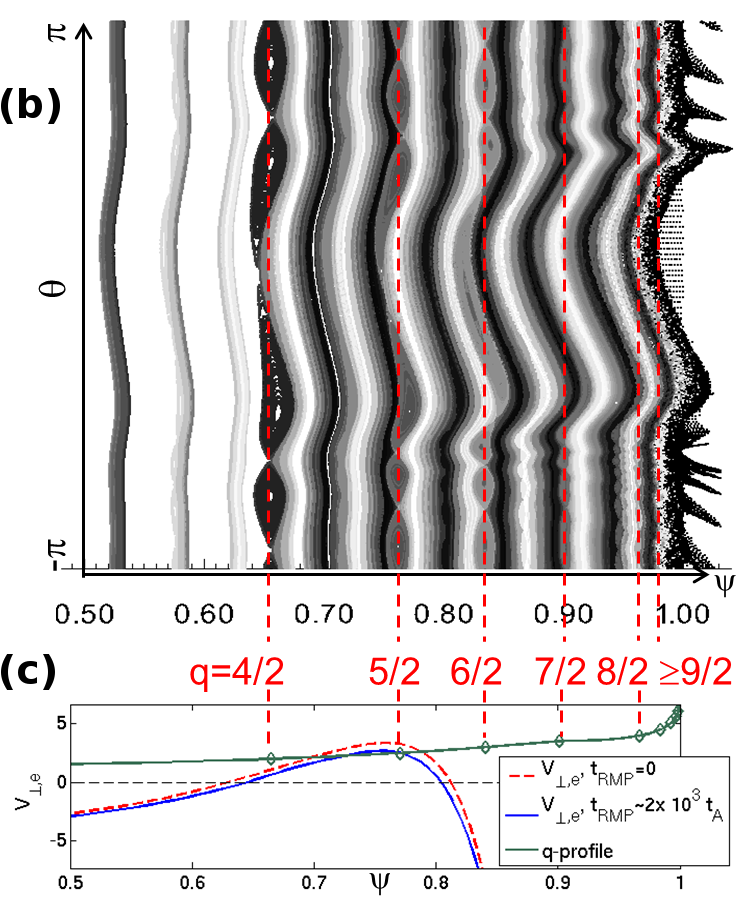} 
\end{minipage}
\centering
\caption{Poincar\'e plot of the magnetic topology as a function of (a) [R,Z] coordinates. (b) [$\psi_{norm}$,$\theta$] coordinates. Resonant surfaces $q=m/n$ are marked in red. (c) Perpendicular electron velocity profile and q-profile as a function of $\psi_{norm}$.}
\label{fig:generic_poinca}
\end{figure}

In all $\Delta \phi$ configurations, the plasma response shows similar features: magnetic island chains are formed on the resonant surfaces $q=4/2$, $5/2$ and $6/2$, as shown in the Poincar\'e plot in ($\psi_{norm}$, $\theta$) coordinates given in Fig.\ref{fig:generic_poinca}(b). This is due to the fact that the perpendicular electron velocity (Fig.\ref{fig:generic_poinca}(c)) is sufficiently small in this region of the plasma to prevent the formation of large screening currents on these resonant surfaces \citep{Nardon_NF10, Becoulet_NF12, Orain_PoP13}. However in the pedestal, the electron perpendicular flow is very large (large radial electric field well), thus screening currents are large on the resonant surfaces $q=7/2$ and $8/2$ and only very small magnetic islands can be seeded on these surfaces. At the very edge (for $q \ge 9/2$ and $\psi_{norm} \ge 0.97$), the resistivity $\eta \propto T^{-3/2}$ is large enough to prevent the screening current formation; subsequently, an ergodic layer is induced by RMP penetration at the very edge. This ergodic layer comes with the formation of lobe structures near the X-point, observable in the Poincar\'e plot in (R,Z) coordinates (Fig.\ref{fig:generic_poinca}(a)) and also observed in experiments \citep{Kirk_PRL12}. 

Even though these general observations are valid for all $\Delta \phi$ configurations, some important characteristics such as the size of the ergodic layer and the lobe structures as well as the kinking of the field lines, differ sensitively depending on $\Delta \phi$, \textit{i.e.} on the applied spectrum, as discussed in Section \ref{sec:part4}.

\section{Impact of differential phase between coils on plasma response} \label{sec:part4}

\begin{figure}[h!]
\centering
\includegraphics[width=1.05\textwidth]{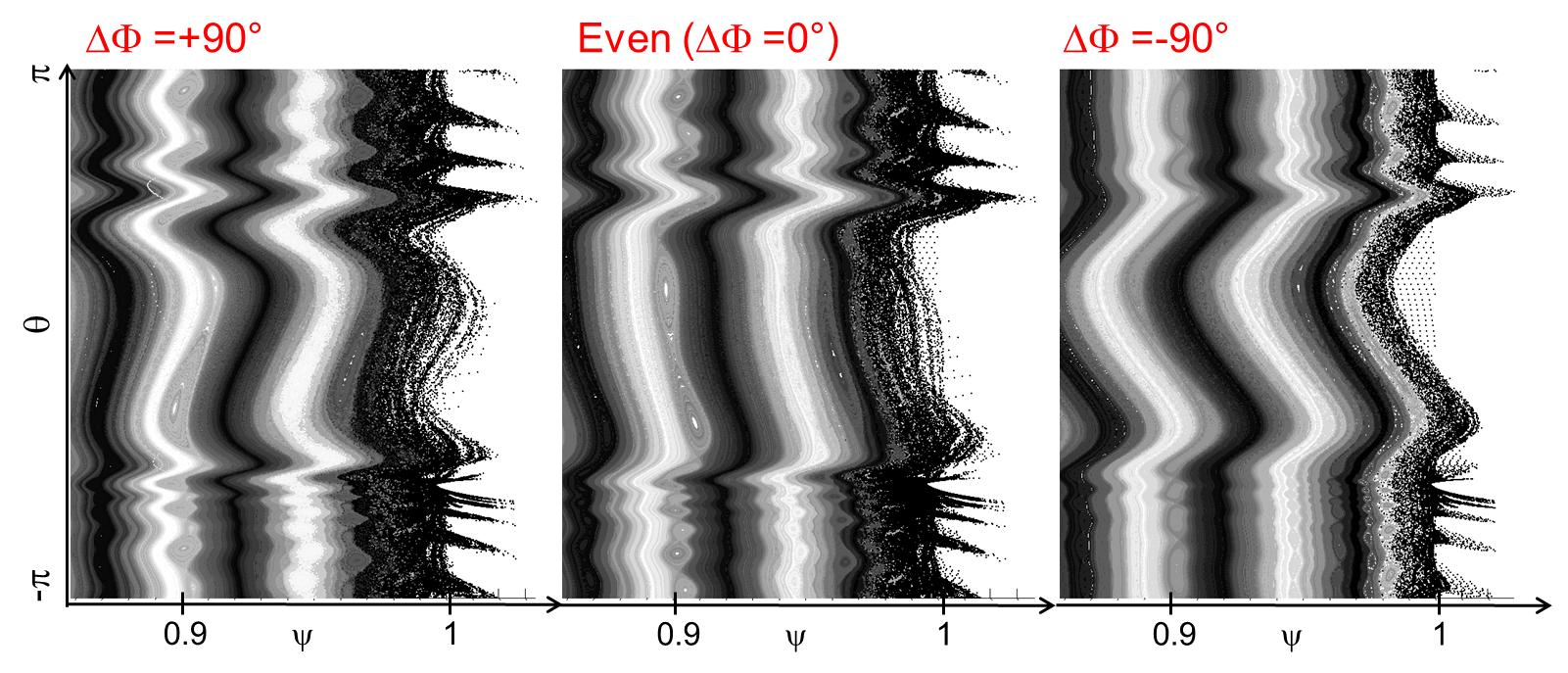}
\caption{Poincar\'e plot of the magnetic topology as a function of  [$\psi_{norm}$,$\theta$] coordinates at the plasma edge ($\psi_{norm} \ge 0.85$) for three RMP configurations: $\Delta \Phi = +90^{\circ}$, $0^{\circ}$ and $-90^{\circ}$.} 
\label{fig:poinca}
\end{figure}

Particularly at the plasma edge, the plasma response substantially depends on the differential phase between the coils applying the RMPs. First, as shown in Fig.\ref{fig:poinca} for three representative simulations, a different kinking of the field line (due to the ideal plasma response -- so called kink response) is observed in the different configurations. The kinking at the edge around the midplane ($\theta = 0$) is maximal for $\Delta \Phi = -90^{\circ}$ and minimal for $\Delta \Phi = 0^{\circ}$, while the kinking near the X-point ($\theta \approx -\pi/2$) is maximal for $\Delta \Phi = +90^{\circ}$ and minimal for $\Delta \Phi = -90^{\circ}$.

\begin{figure}[h!]
\centering
\includegraphics[width=0.5\textwidth]{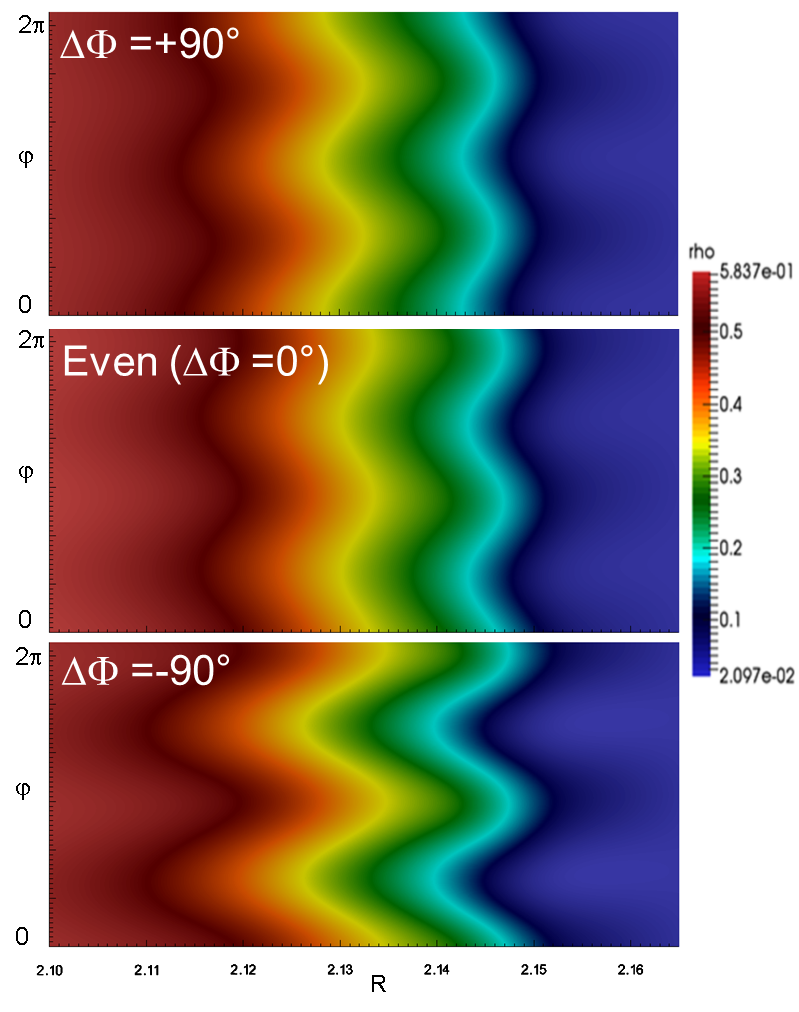}
\caption{Radial profile of the mass density $\rho$ at the edge of the outboard midplane (for $R$ from 2.1 to 2.165) as a function of the toroidal angle $\varphi$, in the three cases $\Delta \Phi = +90^{\circ}$, $0^{\circ}$ and $-90^{\circ}$.} 
\label{fig:disp_midplane}
\end{figure}

Consistently, a similar trend can be observed in the 3D-profiles of density and temperature. At the edge of the outboard midplane, the distortion of the density profile (Fig.\ref{fig:disp_midplane}) when following the toroidal direction is maximal for $\Delta \Phi = -90^{\circ}$ (the radial variation between the maximal and minimal toroidal excursion reaches $\Delta R \approx 0.8mm$ on $q=9/2$) while this displacement is smaller for $\Delta \Phi = 0^{\circ}$ ($\Delta R \approx  4mm$) and $\Delta \Phi = +90^{\circ}$ ($\Delta R \approx  3.5mm$).

\begin{figure}[h!]
\centering
\includegraphics[width=0.75\textwidth]{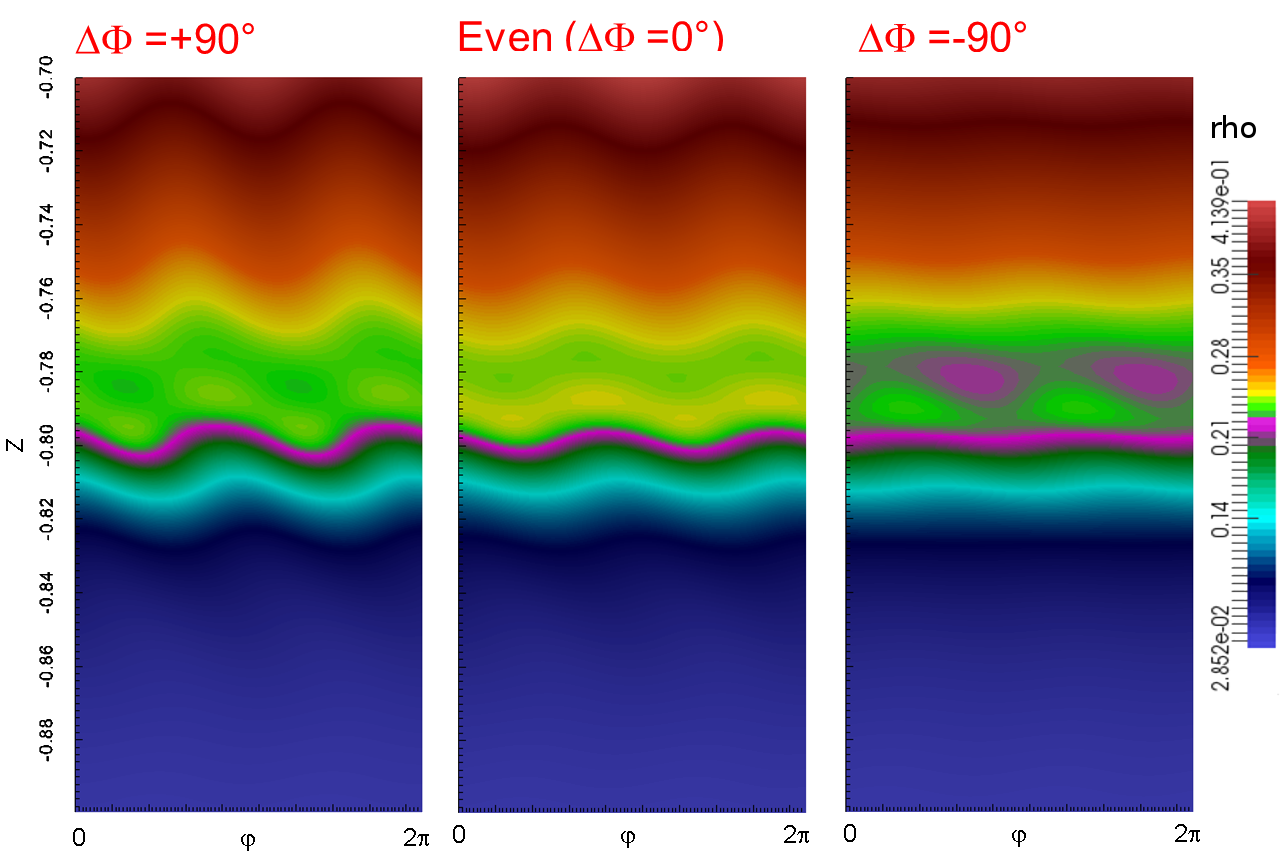}
\caption{Vertical profile of the mass density $\rho$ around the X-point (for $Z$ from -0.7 (over the X-point) to -0.9 (under the X-point), at constant $R=R_{X-point}$) as a function of the toroidal angle $\varphi$, in the three cases $\Delta \Phi = +90^{\circ}$, $0^{\circ}$ and $-90^{\circ}$.} 
\label{fig:disp_Xpoint}
\end{figure}

In the same way, the  distortion of the density profile near the X-point puts in evidence the field line deformation at the vicinity of the X-point. Fig.\ref{fig:disp_Xpoint}, which follows a vertical line passing through the X-point along the toroidal direction, shows that the vertical distortion of the density is largest for $\Delta \Phi = +90^{\circ}$ ($\Delta Z \approx  7mm$ on $q=9/2$, while $\Delta Z \approx  4mm$ for $\Delta \Phi = 0^{\circ}$ and $\Delta Z \approx  2mm$ for $\Delta \Phi = -90^{\circ}$).

It is thus important to notice that the case for which the strongest ELM mitigation is obtained in experiments (largest ELM frequency for $\Delta \Phi = +90^{\circ}$) corresponds to the largest (so-called peeling-kink) displacement near the X-point found in this modeling.

\begin{figure}[h!]
\centering
\includegraphics[width=1.05\textwidth]{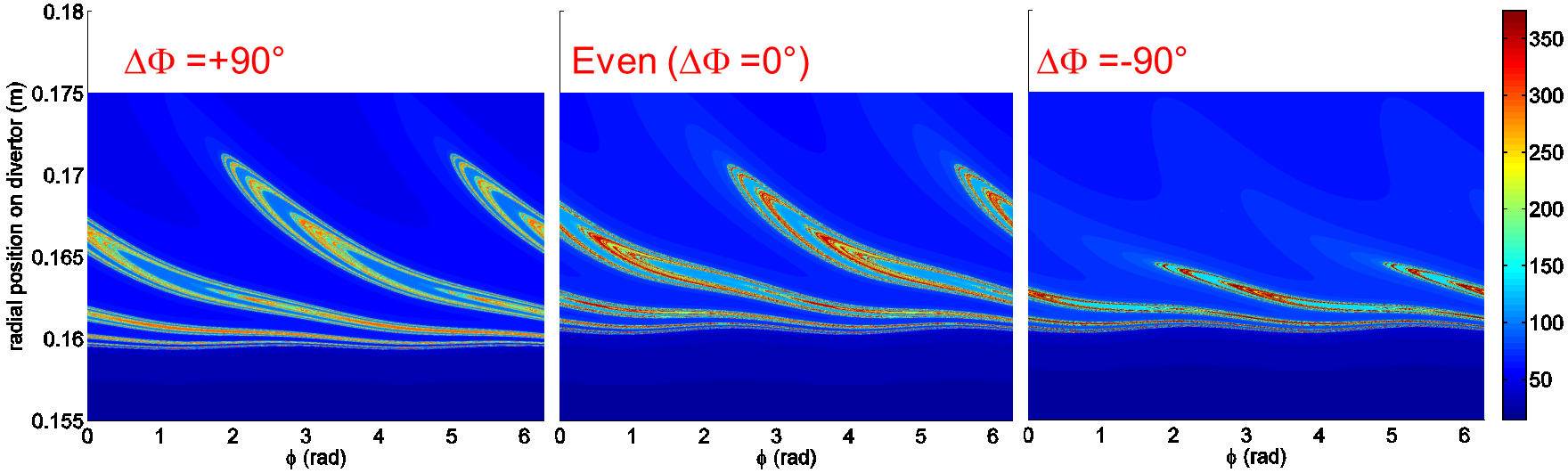}
\caption{Footprints induced on the outer divertor target plate for $\Delta \Phi = +90^{\circ}$, $0^{\circ}$ and $-90^{\circ}$ (connection length plotted as a function of the toroidal angle $\phi$ and the radial position on the divertor target).} 
\label{fig:footprints}
\end{figure}

In addition, this best-ELM-mitigation case in experiments also corresponds to the largest ergodic layer obtained in modeling. Indeed, in Fig.\ref{fig:poinca}, the ergodic layer induced by RMPs at the edge is slightly thicker in the $\Delta \Phi = +90^{\circ}$ case (the magnetic field is ergodic for $\psi_{norm} \ge 0.97$, compared to $\psi_{norm} \ge 0.98-0.99$ in the other cases), and this is correlated with the presence of slightly longer lobe structures near the X-point. The longer lobes also lead to longer footprint ($n=2$) patterns on the divertor, as plotted in Fig.\ref{fig:footprints}: for $\Delta \Phi = +90^{\circ}$, the footprints induced by RMPs on the outer divertor target are slightly larger than for $\Delta \Phi = 0^{\circ}$, and significantly larger than for $\Delta \Phi = -90^{\circ}$. Note that in the $\Delta \Phi = +90^{\circ}$ case, the ($n=2$) magnetic perturbation has non-linearly induced a global ($n=0$) displacement of the X-point, and hence a $2mm$-displacement of the radial position of the footprint on the divertor, observable in Fig.\ref{fig:footprints}. Unfortunately no infra-red thermography data were obtained in this experimental discharge to be compared with the footprints found in modeling.

\begin{figure}[h!]
\centering
\includegraphics[width=0.6\textwidth]{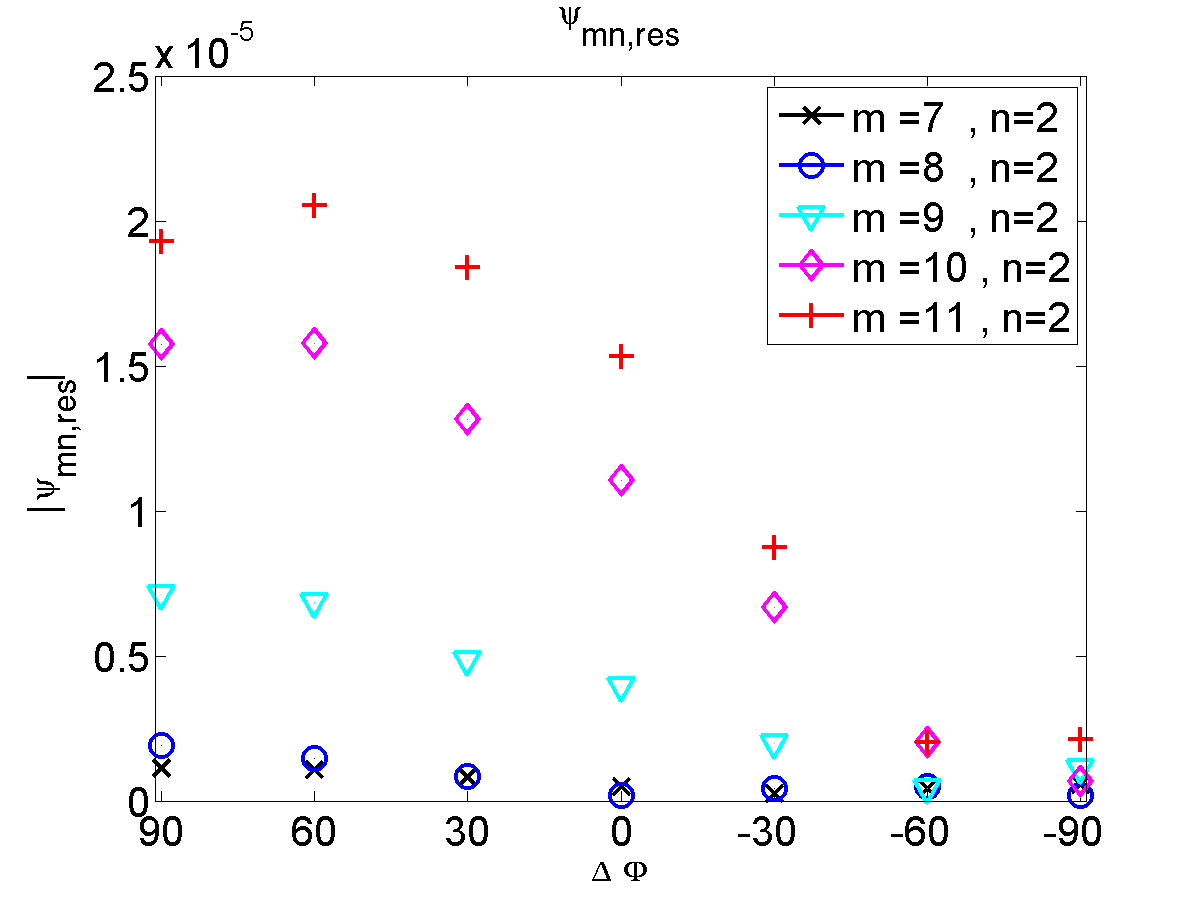}
\caption{$|\psi_{mn,res}|$: $(n=2)$ Fourier harmonics of the magnetic flux perturbation on the resonant surfaces $q=m/n$ for $m=7$ to $11$, plotted as a function of the differential angle between upper and lower RMP coils ($\Delta \phi$).}
\label{fig:psi_mn}
\end{figure}

Furthermore, the size of the ergodic layer (and the size of the lobe structures and footprints as well) is actually correlated with the amplitude of the resonant component of the magnetic perturbation. 
This can be explained by the fact that the resonant components determine the width of the magnetic islands and therefore the width of the overlap region, while being at the same time related to a quantity which determines the lobe length (Poincar\'e \cite{abdullaev:042508} or Melnikov \cite{pavel-eric-footprints} integrals).
Indeed, the amplitudes of the ($n=2$) Fourier harmonic of the magnetic flux $|\psi_{mn,res}|$ on the resonant surfaces $q=m/n$ (plotted in Fig.\ref{fig:psi_mn} as a function of the differential angle between upper and lower RMP coils for $m=7$ to $11$), are maximum for the $\Delta \Phi = +90/+60^{\circ}$ cases. The resonant flux perturbation on $q=7/2$ and $8/2$ is small in all configurations due to the large screening induced by the large perpendicular electron flow at the pedestal top, as previously discussed in Secs. \ref{sec:part2}-\ref{sec:part3}. However, on the resonant surfaces located close to the separatrix ($q=m/2$ for $m \ge 9$), the resonant component is largest for $\Delta \Phi = +90/+60^{\circ}$ and is reduced as $\Delta \Phi$ is reduced, reaching a 10 times lower value for $\Delta \Phi = -60/-90^{\circ}$: this means that the width of the stochastic layer at the edge is progressively reduced as $\Delta \Phi$ is reduced during this discharge (Fig.\ref{fig:phase_scan}).

\begin{figure}[h!]
\centering
\includegraphics[width=1.05\textwidth]{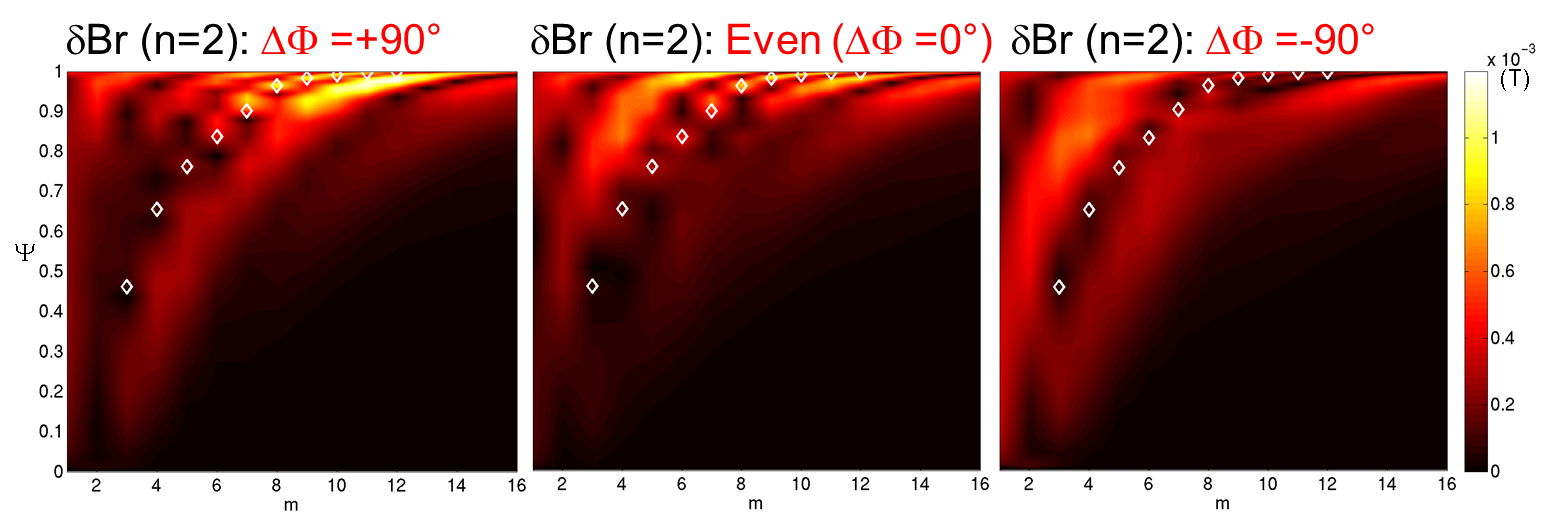}
\caption{$(n=2)$ Radial perturbation of the magnetic field as a function of the poloidal mode number $m$ and the radial direction $\psi_{norm}$, for the three cases: $\Delta \Phi = +90^{\circ}$, $0^{\circ}$ and $-90^{\circ}$. Resonant surfaces are marked with white diamonds.} 
\label{fig:deltaBr}
\end{figure}

The 2D-plot of the ($n=2$) component of the radial magnetic field as a function of the poloidal mode number $m$ and the radial direction $\psi_{norm}$ (Fig.\ref{fig:deltaBr}) highlights particularly well the amplitude of the resonant components (marked by white diamonds) and the kink components, for the different cases $\Delta \Phi = +90^{\circ}$, $0^{\circ}$ and $-90^{\circ}$. In all cases, the resonant component is close to zero at the center due to the strong screening. At the edge (for $\psi_{norm} > 0.95$), the resonant component (on $q=m/2$ with $m \ge 8$) has a large amplitude for $\Delta \Phi = +90^{\circ}$, a smaller amplitude for $\Delta \Phi =0^{\circ}$ and a very small amplitude for $\Delta \Phi = -90^{\circ}$. As for the kink component, composed of modes $m>n q$ \citep{Reimerdes_NF09}, its amplitude is relatively large in the core in the $\Delta \Phi = -90^{\circ}$ and $+90^{\circ}$ cases, but the kink amplitude is especially large at the edge in the $\Delta \Phi = +90^{\circ}$ case. This edge amplification corresponds to the peeling-kink displacement near the X-point previously discussed in this section.

\begin{figure}[h!]
\centering
\includegraphics[width=0.6\textwidth]{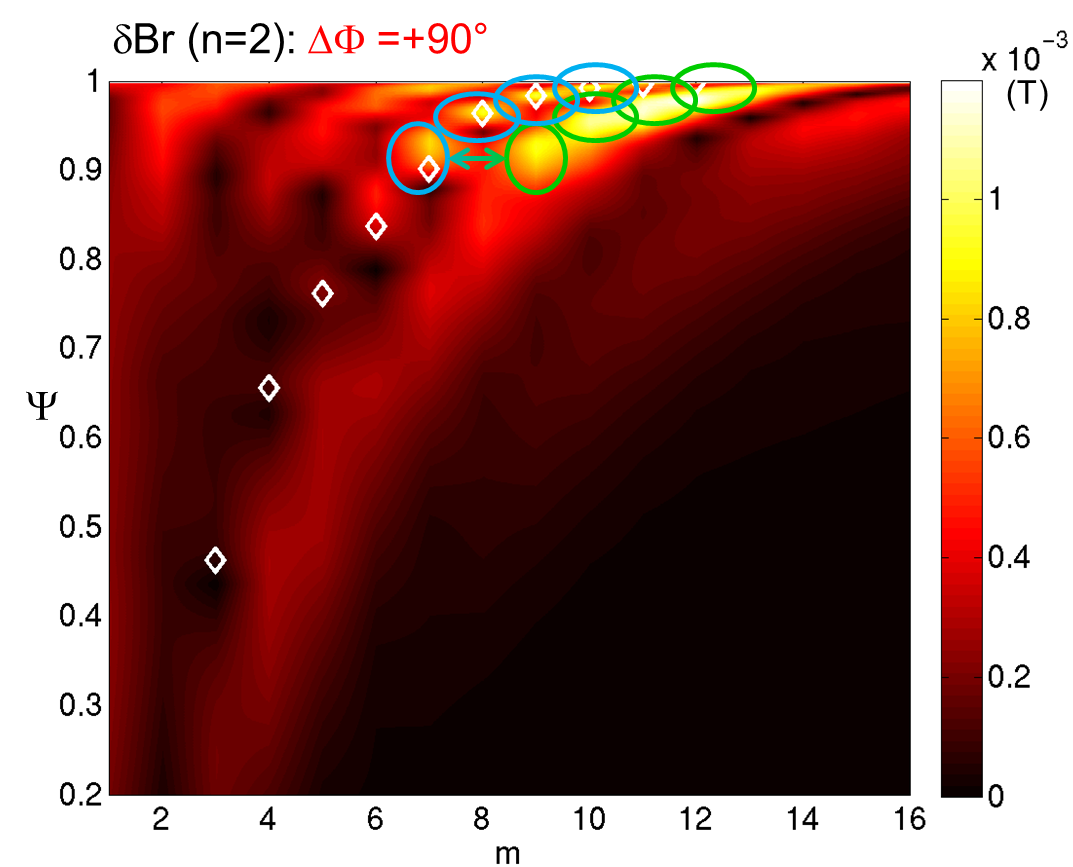}
\caption{Zoom of Fig.\ref{fig:deltaBr} for the $\Delta \Phi = +90^{\circ}$ case, highlighting the coupling between resonant $m$ component with the kink $m+2$ component.} 
\label{fig:deltaBr_zoom}
\end{figure}

A zoom into Fig.\ref{fig:deltaBr} for the $\Delta \Phi = +90^{\circ}$ case (Fig.\ref{fig:deltaBr_zoom}) allows to exhibit the strong coupling between the resonant $m$ component (circled in blue) and the $m+2$ peeling-kink component (circled in green). This suggests that the peeling-kink amplification near the X-point induces the amplification (or prevent the screening) of the resonant magnetic perturbation at the edge. The amplified resonant component therefore induces a larger ergodicity at the edge and an increased radial transport.

\section{Evolution of the background profiles} \label{sec:part5}

\begin{figure}[h!]
\hspace{-0.5cm}
\begin{minipage}[]{0.45\linewidth}
\includegraphics[width=1.\columnwidth]{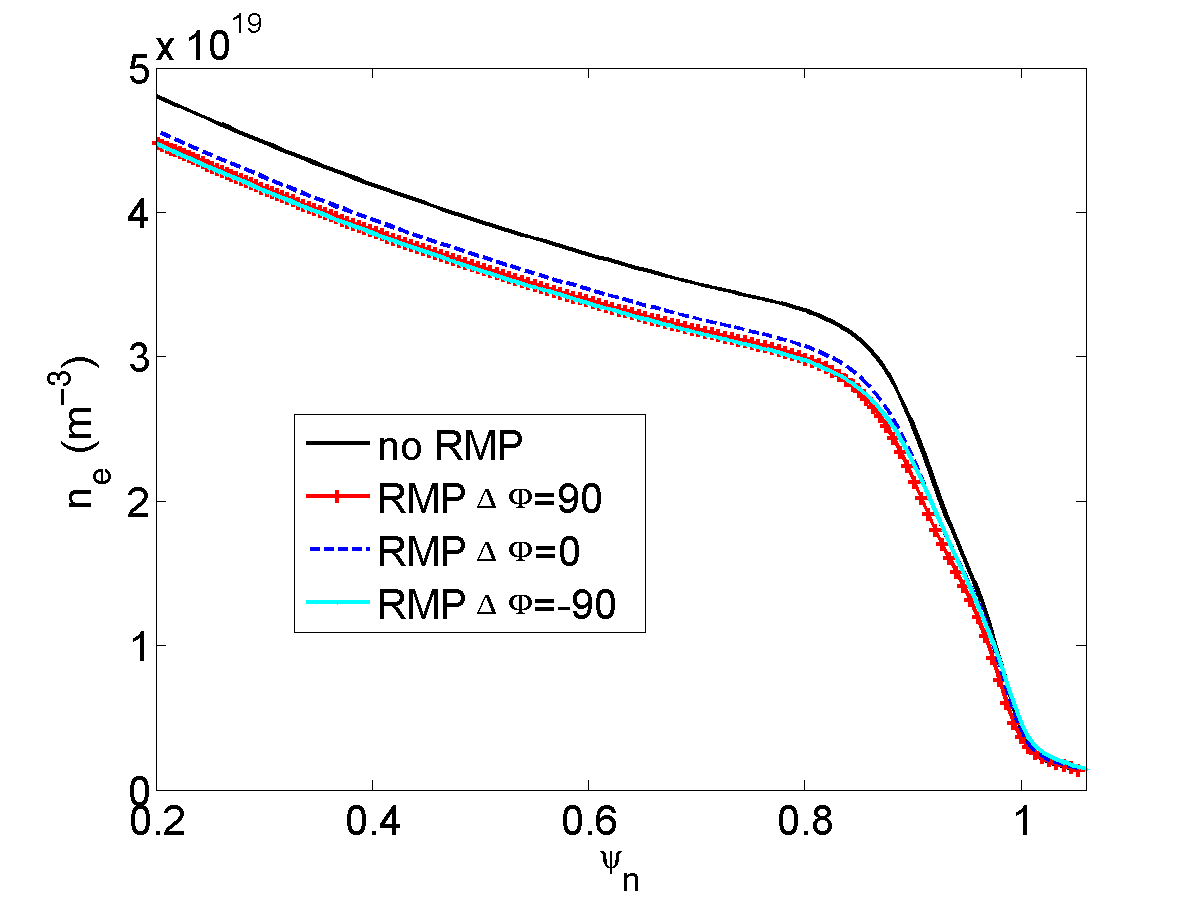}
%
\includegraphics[width=1.\columnwidth]{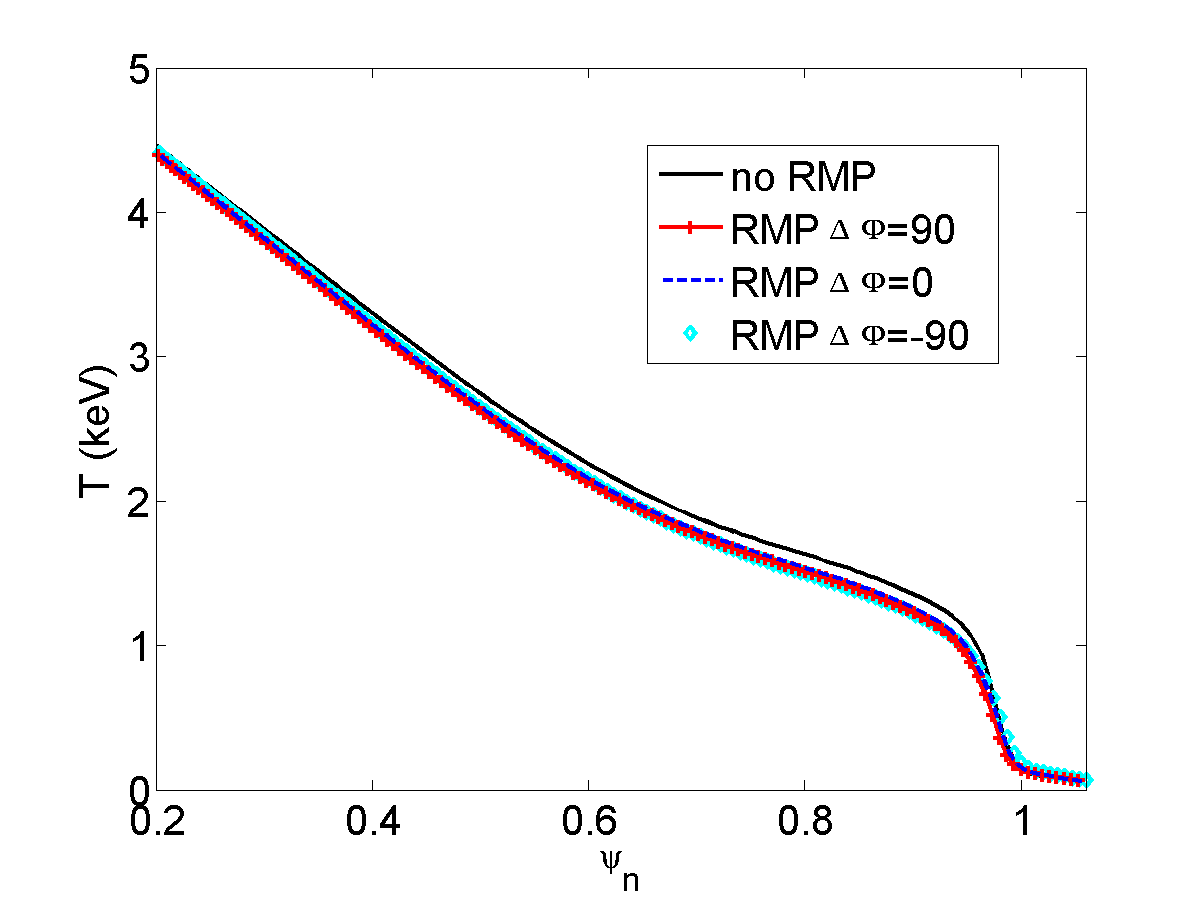} 
\end{minipage}
\begin{minipage}[]{0.45\linewidth}
\includegraphics[width=0.84\columnwidth]{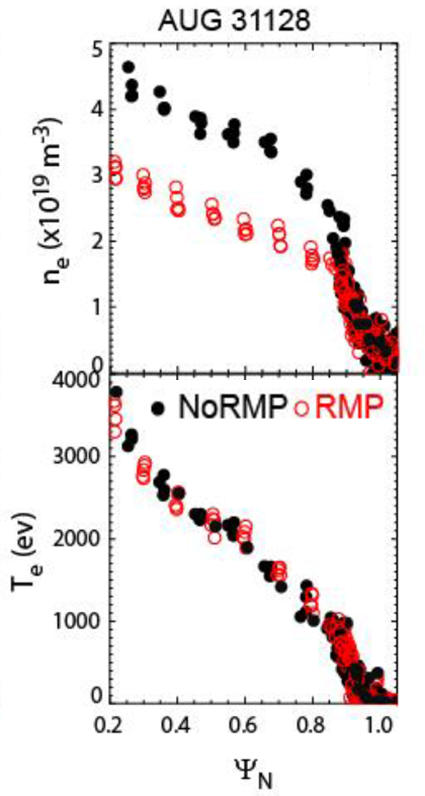} 
\end{minipage}
\centering
\caption{(Left) Radial profiles of electron density $n_e$ and electron temperature $T_e$ at outboard midplane obtained in modeling without RMPs and with RMPs for $\Delta \Phi = +90^{\circ}$, $0^{\circ}$ and $-90^{\circ}$, $60ms$ after the RMP application. (Right) Comparison with experimental profiles without and with RMPs ($\Delta \Phi = +90^{\circ}$ case).}
\label{fig:pumpout}
\end{figure}

The 3D density profile is not only corrugated following the $n=2$ displacement, but the axisymmetric $n=0$ profile is also non-linearly affected by $n=2$ RMPs. Thus a significant reduction of the overall density (so-called density pumpout) is observed in modeling, as shown in Fig.\ref{fig:pumpout}. The outboard midplane density profile is plotted without RMPs and with RMPs in the three configurations $\Delta \Phi = +90^{\circ}$, $0^{\circ}$ and $-90^{\circ}$, $60ms$ after switching on the RMPs. The comparison of the $\Delta \Phi = +90^{\circ}$ case with the experimental profiles (shot \#31128, Fig.\ref{fig:pumpout} (right)) shows that the density pumpout obtained in modeling is actually smaller than the one observed in experiments, but is still significant enough to be underlined. The time scale of pumpout to be fully developed in experiments is around $\sim 200ms$ so the profiles in modeling only show partially  the density reduction. However other additional mechanisms such as the density transport by magnetic flutter \citep{Waelbroek_NF12} and by turbulence \citep{Mordijck_PPCF15}, not included in this model, are good candidates to explain this discrepancy and remain to be studied in future works. Corrections in the radial electric field, as suggested in \cite{Kaveeva_NF12}, might also be considered. 

In addition, the density pumpout observed in modeling is slightly larger in the $\Delta \Phi = +90^{\circ}$ case than for other RMP configurations, but the difference is not significant enough as compared to the difference observed in the experiments. So the additional effects quoted above may be necessary to consistently model the actual density pumpout.

As for the temperature profiles, plotted at the bottom of Fig.\ref{fig:pumpout}, they are barely affected by the application of the RMPs, consistently with the experimental facts. 

\section{Conclusion} \label{sec:part6}

The modeling of the plasma response to RMPs in ASDEX Upgrade, depending on the applied RMP spectrum (changed by varying the differential phase between upper and lower coils), was performed with the resistive non-linear MHD code JOREK. Input profiles and RMP spectrum were chosen to closely match the experimental data. Two-fluid diamagnetic rotation, neoclassical friction and a source of toroidal rotation implemented in the model allowed to reproduce equilibrium rotation profiles similar to the experimental ones. The enhancement of the pedestal density and temperature profiles (in order to make the plasma peeling-ballooning unstable) may nevertheless lead to an overestimated poloidal velocity and thus a slightly overestimated electron perpendicular rotation in the pedestal. Subsequently the shielding of RMPs in the pedestal may be accentuated in the modeling by the large electron perpendicular flow. On the other hand, the resistivity taken in modeling (10 times larger than the Spitzer value) should on the contrary increase the size of magnetic islands and the width of the ergodic layer. In spite of these limitations, the qualitative behaviour of the plasma response to RMPs is modeled consistently with experimental observations.

For all RMP configurations, a similar behaviour was observed in modeling: the screening of magnetic perturbations prevent the formation of significant islands at the pedestal top due to the large electron perpendicular flow. At the very edge (last $1$ to $3\%$ of the normalized poloidal flux inside the separatrix), an ergodic layer is induced by the RMP penetration, due to the local high resistivity (proportional to $T^{-3/2}$). RMP also induce the 3D deformation of the separatrix and the formation of lobe structures near the X-point, which impose an $n=2$ footprint pattern on the divertor target plates.

However significant differences appear depending on the applied RMP spectrum (varied via $\Delta \Phi$). In the $\Delta \Phi = +90^{\circ}$ case, the largest kinking of the field lines is observed near the X-point (as compared to other configurations), also inducing the largest 3D displacement of density and temperature profiles around the X-point. This large peeling-kink amplification near the X-point also generates the amplification of the resonant (tearing) response, via the strong coupling between the $m+2$ peeling-kink component with the $m$ resonant component. Due to the amplified resonant component, a larger ergodic layer is observed in this configuration (also going with longer lobe structures) and a larger density pumpout is non-linearly induced. Even though the density pumpout observed in modeling is not as large as in experiments, it is interesting to note that this configuration (with largest peeling-kink response) corresponds to the strongest ELM mitigation observed in experiments, characterized by the largest increase in ELM frequency and the largest density pumpout. Thus, the amplification of the resonant perturbation, induced by the coupling with peeling-kink modes excited by RMPs, probably generates an increased particle transport near the X-point responsible for ELM mitigation. These findings are in agreement with the results obtained in L-mode experiments and
MARS-F calculations on MAST. As reported in Ref.\cite{Kirk_PPCF11}, the density pump-out in the MAST L-mode experiments was found to be related to the displacement near the X-point calculated by MARS-F, which was also correlated with the amplitude of the resonant components, probably due to a similar coupling mechanism between the peeling-kink mode and the resonant perturbation as described here. The magnetic footprints calculated in the MARS-F field were also longer for the configuration  with a large X-point displacement (in that case, the even parity one), and only in this configuration the footprints were actually experimentally observed \citep{paveliaea2012}.

It is not clear why the ergodicity affects the density profile rather than the temperature profile (which is barely affected by RMPs) but it is consistent with the experimental observations where strong density pumpout is observed while temperature profile remains unchanged \citep{Suttrop_IAEA14, Kirk_NF15}. Note that the width of the ergodic layer is probably too small at the outboard midplane to be observed in experiments, and the displacement near the X-point cannot be measured by diagnostics at this range of parameters. In other experiments at larger magnetic field, the displacement at the midplane could be precisely estimated from ECE diagnostic while rigidly rotating the applied magnetic perturbations \citep{Willensdorfer_PPCF16}, but the displacement near the X-point cannot be determined accurately at the moment. 

Other modeling has been performed using data from the same shots: ideal MHD modeling with VMEC \citep{Suttrop_stellerator15}, resistive linear MHD modeling with MARS-F \citep{Ryan_PPCF15} and linear two-fluid resistive MHD modeling with M3D-$C^1$ \citep{Lyons_APS15}. Good qualitative agreement is found between these simulations and the simulations presented here, on the fact that the strongest ELM mitigation obtained in experiments (for $\Delta \Phi$ between $+90^{\circ}$ and $+60^{\circ}$) is correlated to the largest excitation of the peeling-kink response near the X-point and thus, to the largest displacement around the X-point. The coupling between the peeling-kink component with the resonant component was also found in MARS-F calculations \citep{Ryan_PPCF15}, consistent with the results presented here. 

The density pumpout induced by RMPs is however a feature that can only be reproduced in simulations including the non-linear coupling between the ($n=2$) perturbation and the ($n=0$) background profiles. This work shows that this pumpout can be partially explained by the ergodization and a large 3D-displacement near the X-point, yet some physical ``ingredients'' such as the magnetic flutter (radial particle transport induced by the current flow along the perturbed field lines) will be introduced in the model in the near future in order to improve the modeling of the pumpout. Turbulence may also play a role on the increased particle transport induced by RMPs, which cannot be taken into account in this model.

Present and future work will focus on the non-linear interaction between ELMs and RMPs, aiming to study the different couplings of $n=2$ RMPs with unstable modes, depending on the plasma response. In addition, comparisons with DIII-D cases at low collisionality will be made, since the ELM suppression at low collisionality observed in DIII-D \citep{Paz-Soldan_PRL15, Nazikian_PRL15} is possibly correlated to the excitation of the peeling-kink response near the X-point, similarly to the observations in ASDEX Upgrade.

\footnotesize\section*{Acknowledgement}
This work has been carried out within the framework of the EUROfusion Consortium and has received funding from the Euratom research and training programme 2014-2018 under grant agreement No 633053. Part of this work was carried out using the HELIOS supercomputer system at Computational Situational Centre of International Fusion Energy Research Centre (IFERC-CSC), Aomori, Japan, under the Broader Approach collaboration between Euratom and Japan, implemented by Fusion for Energy and JAEA. The work of Pavel Cahyna was supported by Czech Science Foundation grant 16-24724S and by MSMT CR, grant \#8D15001. The views and opinions expressed herein do not necessarily reflect either those of the European Commission or those of the ITER Organization.

\bibliographystyle{unsrt}
\bibliography{NF_paper_2016_submitted1}

\end{document}